\documentclass[review,number,sort&compress]{elsarticle}

\usepackage{tikz}
\usepackage{float}
\usepackage{graphicx}
\usepackage{gensymb}
\usepackage{amssymb}
\usepackage{color,soul}
\usepackage{url}
\usepackage{ulem}
\usepackage{textcomp}
\usepackage[figuresright]{rotating}
\usepackage{subcaption}
\usepackage{lineno}
\usepackage{hyperref}
\usepackage{cleveref}
\usepackage{libertine}
\usepackage{siunitx}
\usepackage[printwatermark]{xwatermark}
\usepackage[a4paper, total={5.0in, 8in}]{geometry}
\usepackage[T1]{fontenc}

\definecolor{amethystbg}{rgb}{0.6, 0.4, 0.8}
\definecolor{coolgreybg}{rgb}{0.55, 0.57, 0.67}
\definecolor{babypinkbg}{rgb}{0.96, 0.76, 0.76}
\definecolor{cadmiumgreenbg}{rgb}{0.0, 0.42, 0.24}
\definecolor{bluebg}{rgb}{.63,.79,.95}
\definecolor{orangebg}{rgb}{1,0.5,0}
\colorlet{lightbluebg}{bluebg!40}
\colorlet{lightorangebg}{orangebg!40}
\colorlet{lightcadmiumgreenbg}{cadmiumgreenbg!40}
\colorlet{lightbabypinkbg}{babypinkbg!40}
\colorlet{lightcoolgreybg}{coolgreybg!40}
\colorlet{lightamethystbg}{amethystbg!40}

\DeclareGraphicsExtensions{.eps,.pdf,.png,.jpg,}

\journal{Nuclear Instruments and Methods A}

\begin{document}

\newcommand{\etal}{{\it et~al.}}
\newcommand{\geant} {{{G}\texttt{\scriptsize{EANT}}4}}
\newcommand{\srim} {\texttt{SRIM}}
\newcommand{\python} {\texttt{Python}}
\newcommand{\pandas} {\texttt{pandas}}
\newcommand{\SciPy} {\texttt{SciPy}}
\newcommand{\ROOT} {\texttt{ROOT}}

\DeclareRobustCommand{\hlb}[1]{{\sethlcolor{lightbluebg}\hl{#1}}}
\DeclareRobustCommand{\hlo}[1]{{\sethlcolor{lightorangebg}\hl{#1}}}
\DeclareRobustCommand{\hlg}[1]{{\sethlcolor{lightcadmiumgreenbg}\hl{#1}}}
\DeclareRobustCommand{\hlp}[1]{{\sethlcolor{lightbabypinkbg}\hl{#1}}}
\DeclareRobustCommand{\hlgr}[1]{{\sethlcolor{lightcoolgreybg}\hl{#1}}}
\DeclareRobustCommand{\hla}[1]{{\sethlcolor{lightamethystbg}\hl{#1}}}

\begin{frontmatter}


\title{Response of a Li-glass/multi-anode photomultiplier detector to collimated thermal-neutron beams\tnoteref{label1}}



	\author[lund]{E.~Rofors}
	\author[lund]{N.~Mauritzson}
	\author[lund,ess]{H.~Perrey}
	\author[ess,glasgow]{R.~Al~Jebali}
	\author[glasgow]{J.R.M.~Annand}
	\author[glasgow]{L.~Boyd}
        \author[dmsc]{M.J.~Christensen}
	\author[julich]{U.~Clemens}
	\author[LLB]{S.~Desert}
	\author[julich]{R.~Engels}
	\author[lund,ess]{K.G.~Fissum\corref{cor1}}
	\ead{kevin.fissum@nuclear.lu.se}
	\author[julich]{H.~Frielinghaus}
	\author[ideas]{C.~Gheorghe}
	\author[ess,glasgow,milan]{R.~Hall-Wilton}
	\author[julich]{S.~Jaksch}
	\author[ess]{K.~Kanaki}
	\author[ife]{S.~Kazi}
	\author[julich]{G.~Kemmerling}
	\author[ife]{I.~Llamas Jansa}
	\author[lund,ess]{V.~Maulerova}
	\author[glasgow]{R.~Montgomery}
	\author[dmsc]{T.~Richter}
	\author[lund,ess]{J.~Scherzinger\fnref{fn2}}
	\author[glasgow]{B.~Seitz}
	\author[dmsc]{M.~Shetty}

	\address[lund]{Division of Nuclear Physics, Lund University, SE-221 00 Lund, Sweden}
	\address[ess]{Detector Group, European Spallation Source ERIC, SE-221 00 Lund, Sweden}
	\address[glasgow]{SUPA School of Physics and Astronomy, University of Glasgow, Glasgow G12 8QQ, Scotland, UK}
	\address[dmsc]{Data Management and Software Centre, European Spallation Source, Ole Maaløes Vej 3, 2200 Copenhagen, Denmark}
	\address[LLB]{LLB, CEA, CNRS, Universit\'e Paris-Saclay, 91191 Gif-sur-Yvette, France}
	\address[julich]{J\"ulich Centre for Neutron Science JCNS, Forschungszentrum J\"ulich, D-52425 J\"ulich, Germany}
	\address[ideas]{Integrated Detector Electronics AS, Gjerdrums Vei 19, N-0484 Oslo, Norway}
	\address[milan]{Dipartimento di Fisica ``G. Occhialini'', Universit\`a degli Studi di Milano-Bicocca, Piazza della Scienza 3, 20126 Milano, Italy}
	\address[ife]{Institute for Energy Techology, Instituttveien 18, 2007 Kjeller, Norway}

	\tnotetext[label1]{The data set doi:10.5281/zenodo.4095210 is available for download from \href{https://zenodo.org/record/4095210}{https://zenodo.org/record/4095210.}}
	\cortext[cor1]{Corresponding author. Telephone:  +46 46 222 9677; Fax:  +46 46 222 4709}
	\fntext[fn2]{present address: Thermo Fisher Scientific Messtechnik GmbH, Frauenauracher Str. 96, 91056 Erlangen, Germany}

\begin{abstract}
        The response of a position-sensitive Li-glass scintillator detector being 
	developed for thermal-neutron detection with 6\,mm position resolution has 
	been investigated using collimated beams of thermal neutrons. The detector 
	was moved perpendicularly through the neutron beams in 0.5 to 
	1.0\,mm horizontal
	and vertical steps. Scintillation was detected in an 8~$\times$~8 pixel 
	multi-anode photomultiplier tube on an event-by-event basis. In general, 
	several pixels registered large signals at each neutron-beam location. 
	The number of pixels registering signal above a set threshold was 
	investigated, with the maximization of the single-hit efficiency over the 
	largest possible area of the detector as the primary goal. At a threshold 
	of $\sim$50\% of the mean of the full-deposition peak, $\sim$80\% of the 
	events were registered in a single pixel, resulting in an effective 
	position resolution of $\sim$5\,mm in X and Y. Lower thresholds generally
	resulted in events demonstrating higher pixel multiplicities, but these 
	events could also be localized with $\sim$5\,mm position resolution.
\end{abstract}

\begin{keyword}
	SoNDe thermal-neutron detector, GS20 scintillator, Li-glass, H12700A 
	multi-anode photomultiplier, position-dependent response, neutrons, 
	European Spallation Source
\end{keyword}

\end{frontmatter}

\section{Introduction}
\label{section:introduction}

The scientific program to be performed at the European Spallation Source 
(ESS)~\cite{lindroos11,esstdr,garoby18,ess} requires position-sensitive, 
$^{3}$He-free~\cite{kouzes09,shea10,zeitelhack12,kirstein14}, thermal-neutron 
detectors with high counting-rate capability. Small-angle neutron-scattering 
experiments requiring two-dimensional position 
sensitivity~\cite{heiderich91,ralf97,ralf98,ralf99,kemmerling01,ralf02,kemmerling04a,kemmerling04b,jaksch14,feoktystov15}
will be performed with \underline{S}olid-state \underline{N}eutron 
\underline{D}etectors (SoNDe)~\cite{SoNDePatent,sonde,jaksch17cumulative,jaksch18}. 
The SoNDe concept 
employs an array of detector modules to instrument large areas with a 
reconstruction accuracy of $\sim$6\,mm on the position of the detected neutron. 
A SoNDe ``module" consists of a thin, thermal-neutron sensitive, Li-glass 
scintillator sheet (GS20) attached to a 64-pixel multi-anode photomultiplier 
tube (MAPMT). Signals resulting from the scintillation light are processed using 
custom electronics~\cite{jaksch18}. 
In the envisioned operation mode at ESS, known as ``Time of flight" mode (TOF), 
these electronics timestamp all pixels having signals above threshold if any 
single pixel amplitude exceeds threshold. Events involving the firing of a single 
pixel (multiplicity $M$~$=$~1 events) are thus straightforward to interpret. At 
the boundaries between pixels and in the corners, scintillation light sufficient 
to trigger several pixels (multiplicity $M$~$>$~1 events) is often registered. 
The behavior of 
clusters of bordering pixels in these regions is thus of interest. LEDs and laser 
light have been used extensively to study the detailed responses of several 
different 
MAPMTs~\cite{korpar00,rielage01,matsumoto04,lang05,abbon08,montgomery12,montgomery13,calvi15,montgomery15,wang16}.
Previously, scans of a $\sim$1\,mm collimated beam of $\sim$4\,MeV 
$\alpha$-particles from an $^{241}$Am source~\cite{rofors19} and
$\sim$\SI{100}{\micro\meter} diameter beams of 2.5\,MeV protons and 
deuterons~\cite{rofors20} have been used to study the position-dependent response 
of a SoNDe detector prototype in regions well-removed from the edges of the 
detector acceptance. Thermal neutrons have also been used to perform 
first tests both on similar detectors~\cite{zaiwei12} and on SoNDe 
modules~\cite{jaksch18}. The thermal-neutron interaction with the $^{6}$Li of 
the Li-glass has a $Q$-value of 4.78\,MeV and results in an $\alpha$-particle 
(2.05\,MeV) and a triton (2.73\,MeV). In this work, a SoNDe module has been 
systematically scanned through beams of 
thermal~\footnote{	
	The 2.0 and \SI{2.4}{\angstrom} neutron beams employed in this work 
	had energies lying at the upper end of the cold-neutron energy window 
	and are thus ``not quite but nearly" thermal.
	} 
neutrons. The goals 
were to:
\begin{enumerate}
        \item complement the existing $\alpha$-particle, proton, and deuteron
		studies of the position-sensitive behavior of the detector, 
		for events triggering multiple pixels to establish the response
		at the pixel boundaries and the corners where four pixels meet
        \item provide thermal-neutron data with $\sim$1\,mm precision on the 
		position sensitivity of the detector for events triggering only 
		one pixel, as a single-pixel mode-of-operation is anticipated 
		as the ESS default
        \item map the response of the detector as a function of both threshold 
		and beam position for events which only trigger one pixel
        \item determine the detector threshold that maximizes the number of 
		single-pixel events
        \item study regions within the detector where the 
		position-reconstruction accuracy for an event better than 
		$\sim$6\,mm may be obtained for $M$~$>$~1 events
	\item provide a thermal-neutron dataset at the edge of the detector,
		for adjacent pixels with the highest gain contrast, to aid our 
		understanding of the SoNDe module at its periphery
	\item implement a SoNDe module into a prototype of the ESS data-acquisition 
		architecture.
\end{enumerate}

\section{Apparatus}
\label{section:apparatus}

\subsection{Neutron beams}
\label{subsection:NeutronBeams}

\begin{figure}[H]
        \begin{center}
                \begin{subfigure}[b]{0.78\textwidth}
                        \includegraphics[width=1.0\textwidth]{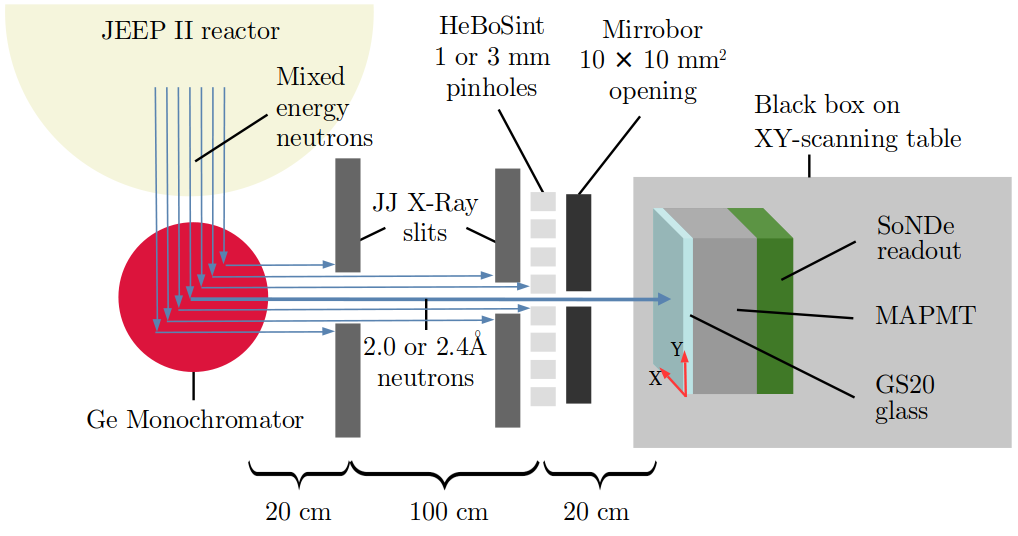}
			\caption{Beamline schematic (not to scale).}
                        \vspace*{4mm}
                        \label{subfig:IFE_overview}
                \end{subfigure}
                \begin{subfigure}[b]{0.68\textwidth}
                        \includegraphics[width=1\textwidth]{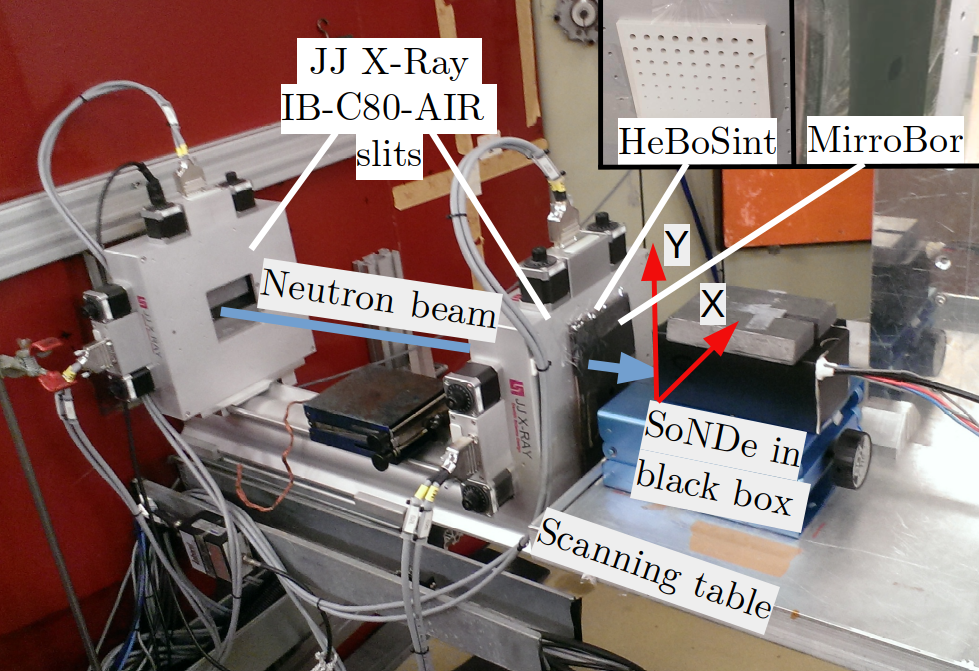}
			\caption{Beamline photo.}
                        \label{subfig:IFE_photo}
                \end{subfigure}
                \caption{
			R2D2 thermal-neutron beamline at the JEEP~II reactior at 
			IFE.
			\ref{subfig:IFE_overview}: The reactor (yellow half circle, 
			top left) produced a continuous beam of neutrons (blue 
			arrows) which was collimated and monochromated (red 
			circle). The beam was then shaped with a pair of slits, a 
			pinhole mask, and a shielding sheet before striking a black 
			box containing the SoNDe module from the left.
			\ref{subfig:IFE_photo}: A photograph of the experimental 
			setup. The neutron beam is shown by the blue arrow from the
			left. The insets (top right) show the pinhole mask and the 
			shielding sheet. The SoNDe module was located in the black 
			box which was mounted on a motorized XY scanning table.
			(For interpretation of the references to color in this
                        figure caption, the reader is referred to the web version
                        of this article.)
			}
        	\label{figure:IFE}
        \end{center}
\end{figure}

The measurements were performed at the R2D2 beamline at the JEEP~II 
reactor~\cite{jeep2,hauback00} at the Institute for Energy Technology 
(IFE)~\cite{IFE} in Norway. The setup is illustrated in Fig.~\ref{figure:IFE}.
The nominal central-beam 
flux was 10$^5$/s/cm$^2$ with a $\sim$0.6$^{\circ}$ divergence. Thermal-neutron 
beams (\SI{2.0}{\angstrom}, $\sim$18\,meV and \SI{2.4}{\angstrom}, $\sim$13\,meV) 
were defined using a composite Ge wafer monochromator~\cite{hauback00}. The 
resulting thermal-neutron beams drifted $\sim$20\,cm to the first of a pair of JJ 
X-Ray IB-C80-AIR slits~\cite{jjxray} which employed borated-aluminum blades to 
control the beam flux. The slit spacing was $\sim$100\,cm, with the downstream 
slit located $\sim$20\,cm upstream of the detector. A 5\,mm thick HeBoSint 
mask~\cite{hebosint} with pinholes was used to further collimate the beam to 
either $\sim$1\,mm or $\sim$3\,mm in diameter. A stack of three 2\,mm thick 
Mirrobor sheets~\cite{mirrobor} with a 100\,mm$^{2}$ square aperture acted 
as a final barrier to any neutrons surviving the upstream collimation. The 
resulting beam was incident upon a black box containing the SoNDe module under 
investigation.

\subsection{SoNDe module}
\label{subsection:detector}

\begin{figure}[H]
	\begin{center}
		\begin{subfigure}[b]{0.32\textwidth}
                       \includegraphics[width=1.0\textwidth]{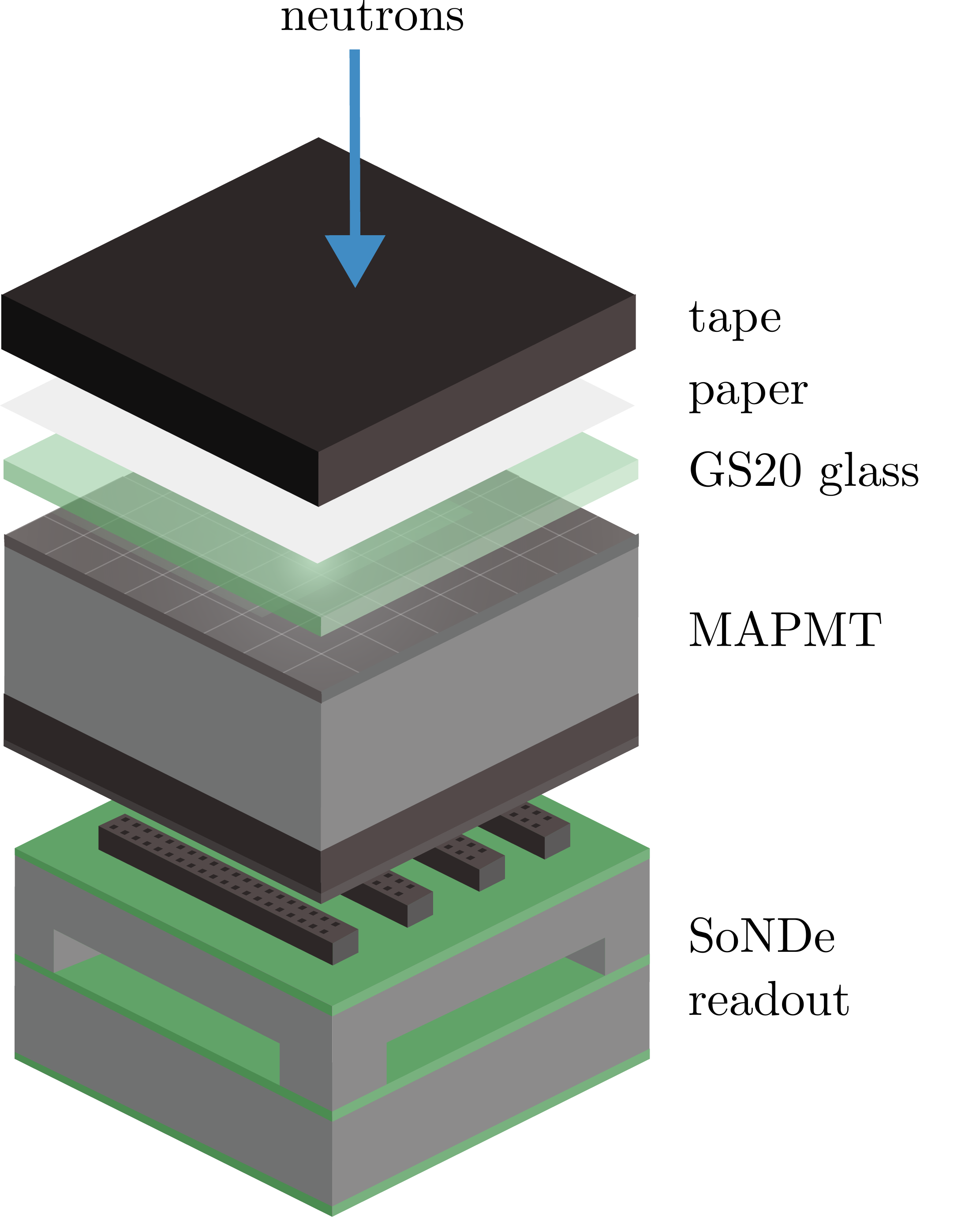}
                       \vspace*{3mm}
                       \caption{SoNDe module \linebreak (exploded view)}
                       \label{subfig:schematic}
		\end{subfigure}
		\begin{subfigure}[b]{0.32\textwidth}
                       \includegraphics[width=1\textwidth]{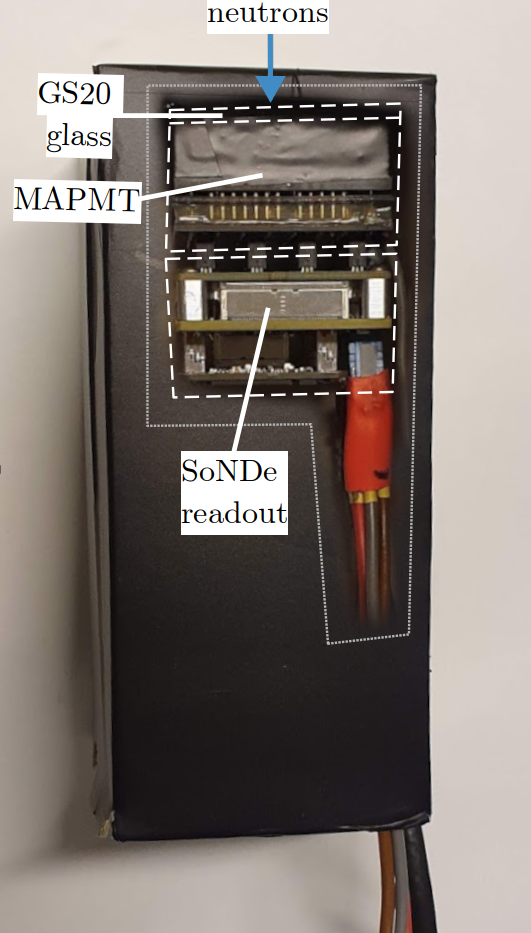}
                       \vspace*{0mm}
                       \caption{SoNDe module inside black box\linebreak (top view)}
                       \label{subfig:photograph}
		\end{subfigure}
		\begin{subfigure}[b]{0.32\textwidth}
			\includegraphics[width=1.\textwidth]{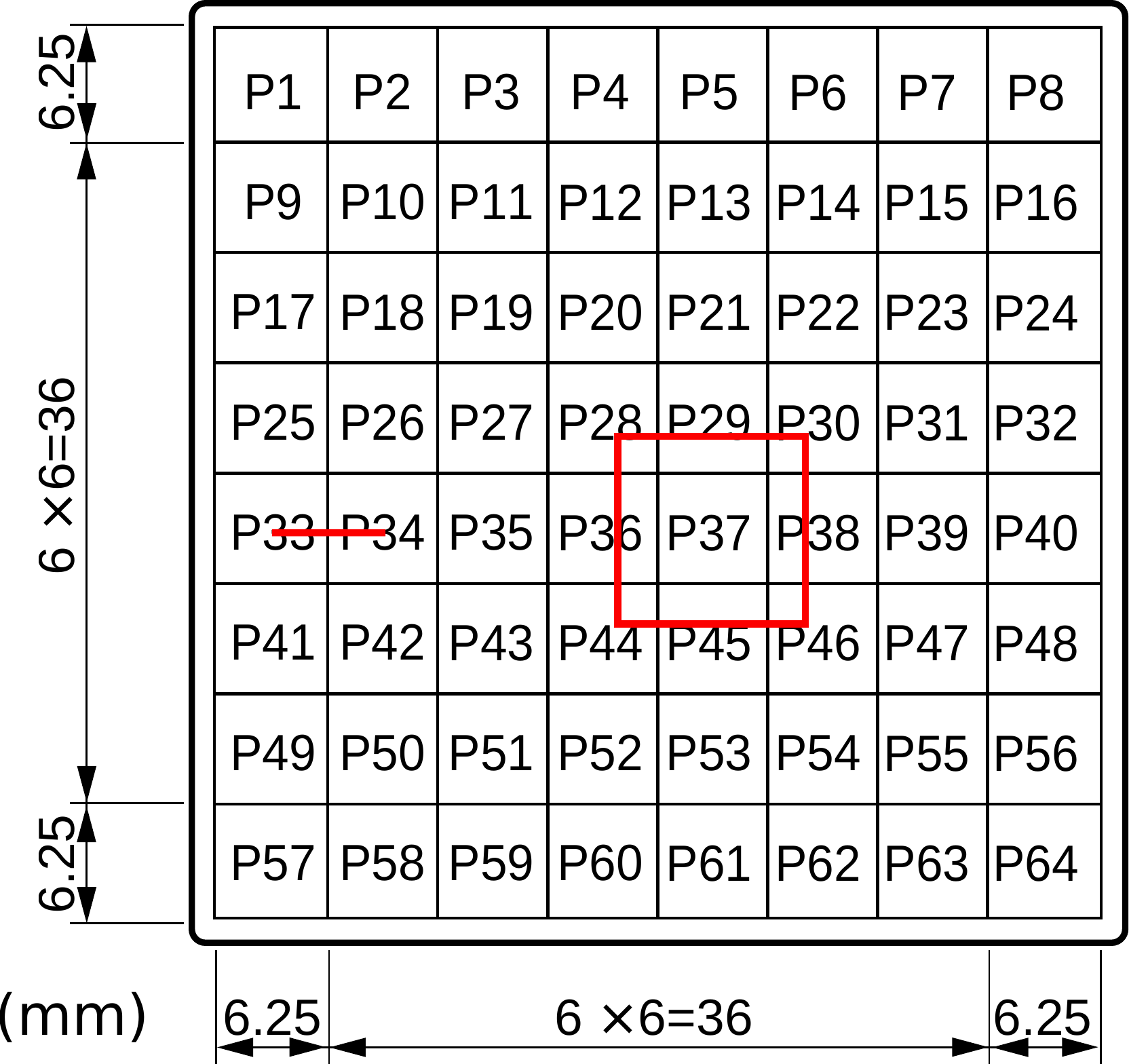}
			\vspace*{7mm}
			\caption{MAPMT pixel map \linebreak (front view)}
			\label{subfig:pixelmap}
		\end{subfigure}
                \caption{
			A SoNDe module.  
			\ref{subfig:schematic}: 3D view of a SoNDe module. From 
			the top: light-tight tape, paper reflector, GS20 glass, 
			MAPMT, and readout electronics. The beam of neutrons (blue 
			arrow) is incident from the top.
			\ref{subfig:photograph}: Photograph of the SoNDe module 
			(dashed white box) in the black box. The GS20 glass, 
			MAPMT, and readout electronics are labeled. The beam of 
			neutrons (blue arrow) is incident from the top.
			\ref{subfig:pixelmap}: Numbering scheme for the MAPMT 
			pixels (front view)~\cite{hamabrochure}. The region of 
			irradiation with the 3\,mm FWHM beam is indicated by the 
			red box centered on P37. The trajectory of irradiation 
			with the 1\,mm FWHM beam is indicated by the red line 
			connecting the center of edge pixel P33 to the center of 
			P34. 
			The beam of neutrons is incident into the page.
			(For interpretation of the references to color in this 
			figure caption, the reader is referred to the web version 
			of this article.)
	       		}
		\label{figure:SoNDeModule}
	\end{center}
\end{figure}

A SoNDe module (Fig.~\ref{figure:SoNDeModule}) described in the following sections 
consists of three basic components:
\begin{enumerate}
	\item a 1\,mm thick Li-glass scintillator sheet
        \item a H12700A MAPMT
        \item purpose-built SoNDe readout electronics.
\end{enumerate}

\subsubsection{Li-glass scintillator}
\label{subsubsection:Liglass_scintillator}

The scintillator employed was a cerium-activated lithium-silicate glass known as 
GS20~\cite{firk61,spowart76,spowart77,fairly78}. Provided by 
Scintacor~\cite{scintacor}, it was developed for the detection of thermal 
neutrons.  The 50\,mm~$\times$~50\,mm~$\times$~1\,mm sheet with polished faces 
and rough-cut 1\,mm edges was dry-fitted to the MAPMT window and held in place 
with tape along the edges. The dry-fit approach was chosen to avoid the 
degradation of any optical-coupling medium. A piece of 
standard white copy paper (136\,g/cm$^{2}$) placed over the upstream face of the 
GS20 diffusely reflected scintillation light back towards the MAPMT, increasing
the amount of scintillation light reaching the MAPMT by $\sim$40\%. The 
(assumed uniform) density of~~$^{6}$Li in GS20 is 
1.58~$\times$~$10^{22}$~atoms/cm$^{3}$. The cross section for the n (25\,meV) + 
$^{6}$Li $\rightarrow$ $^{3}$H (2.73\,MeV) + $\alpha$ (2.05\,MeV) capture 
reaction is 940~b, which yields a detection efficiency of $\sim$75\% for the 
1\,mm sheet. The average ranges of the $^{3}$H and $\alpha$-particle in the GS20 
are \SI{34.7}{\micro\meter} and \SI{5.3}{\micro\meter}, 
respectively~\cite{jamieson15}. The 4.78\,MeV capture reaction results in a
$\sim$6600 scintillation photon full-deposition peak~\cite{jaksch18} (roughly
equivalent to 20-30\% of anthracene) peaked at $\sim$390\,nm~\cite{vanEijk04}. 
Scintillation-light transport from the GS20 (refractive index~1.55 at 395\,nm) 
across a $\sim$\SI{100}{\micro\meter} air gap (refractive index~1) due to the 
concavity of the MAPMT borosilicate-glass window and then into the MAPMT window 
(refractive index~1.53) is generally inefficient.

\subsubsection{Multi-anode photomultiplier tube}
\label{subsubsection:mapmt}

An 8~$\times$~8 pixel ($\sim$6\,mm~$\times$~$\sim$6\,mm per pixel) 10 stage 
Hamamatsu type H12700A MAPMT with a borosilicate glass window was employed.
The sensitive bialkali photocathode area is 48.5\,mm~$\times$~48.5\,mm, while 
the outer dimensions are 52~mm~$\times$~52~mm, resulting in 87\% of the surface 
being active. The peak quantum efficiency of the photocathode is $\sim$33\% at 
$\sim$380\,nm, nicely overlapping the $\sim$390\,nm (peak) scintillation-light 
distribution produced by GS20. For the MAPMT used here, at a cathode-to-anode 
voltage of $-$1000\,V, the Hamamatsu data sheet specified a gain of
2.09~$\times$~10$^{6}$, a dark current of 2.67\,nA, a (worst-case) factor 
1.7 gain difference between pixels, and $\sim$2\,\% electronic crosstalk between 
pixels. Calvi~\etal~\cite{calvi15} report that electronic crosstalk is actually 
dependent upon both the pixel and the position within this pixel and that it 
fluctuates differently near horizontal and vertical edges (see below). The 
operating voltage was -900\,V. Corrections for pixel-to-pixel gain variations 
were performed offline using a gain map measured with the 3\,mm, 
\SI{2.4}{\angstrom} neutron beam used to irradiate each pixel center 
consecutively, as described in \cite{rofors20}.

\subsubsection{Readout electronics}
\label{subsubsection:readout}

A compact readout module designed for the H12700A MAPMT by IDEAS~\cite{ideas} 
was employed for data acquisition. The 113\,g module is 
50~mm~$\times$~50~mm~$\times$~55~mm (deep). It consists of two boards: front-end 
and controller~\cite{jaksch18}. The front-end board uses four 16-channel IDE3465 
ASICs~\cite{meier13} to digitize the MAPMT signals with a precision of 14~bits. 
The controller board accomodates an FPGA and a MiniIO port for ethernet 
communication. The electronics can operate in two modes: the TOF mode previously
discussed and ``All-channel Spectroscopy" (ACS) that was used here. 
In ACS mode, when any pixel-amplitude threshold was exceeded, the digitized 
signal amplitudes from all 64 pixels were read out. The ACS rate limitation is 
$\sim$10\,kHz, which corresponds to $\sim$4\,MHz/m$^{2}$. A hardware threshold of 500~ADC 
channels was employed, which corresponds to $\sim$5\% of the mean channel of 
the 4.78\,MeV full-energy deposition peak measured by a single pixel for 
irradiation at its center. Higher thresholds were applied offline.
Control, visualization, and data logging were provided by the ESS Event
Formation Unit (EFU)~\cite{EFU, EFU2, EFU3} and the ESS Daquiri visualization
tool~\cite{daquiri} both running on a Centos~7 PC. The SoNDe module and the EFU
were connected via switched 1 Gbit/s Ethernet. The SoNDe module is configured
with TCP/IP and transmits readout data over UDP/IP to the EFU~\cite{christensen17}
in a manner similar to that anticipated for operation at ESS. The EFU, designed
for use by ESS instruments, employs an acquisition that also closely resembles
the anticipated operation mode for ESS.

\subsection{\geant~simulation}
\label{subsection:simulation}

\begin{figure}[H]
	\begin{center}
		\begin{subfigure}[b]{0.52\textwidth}
			\includegraphics[width=1.0\textwidth]{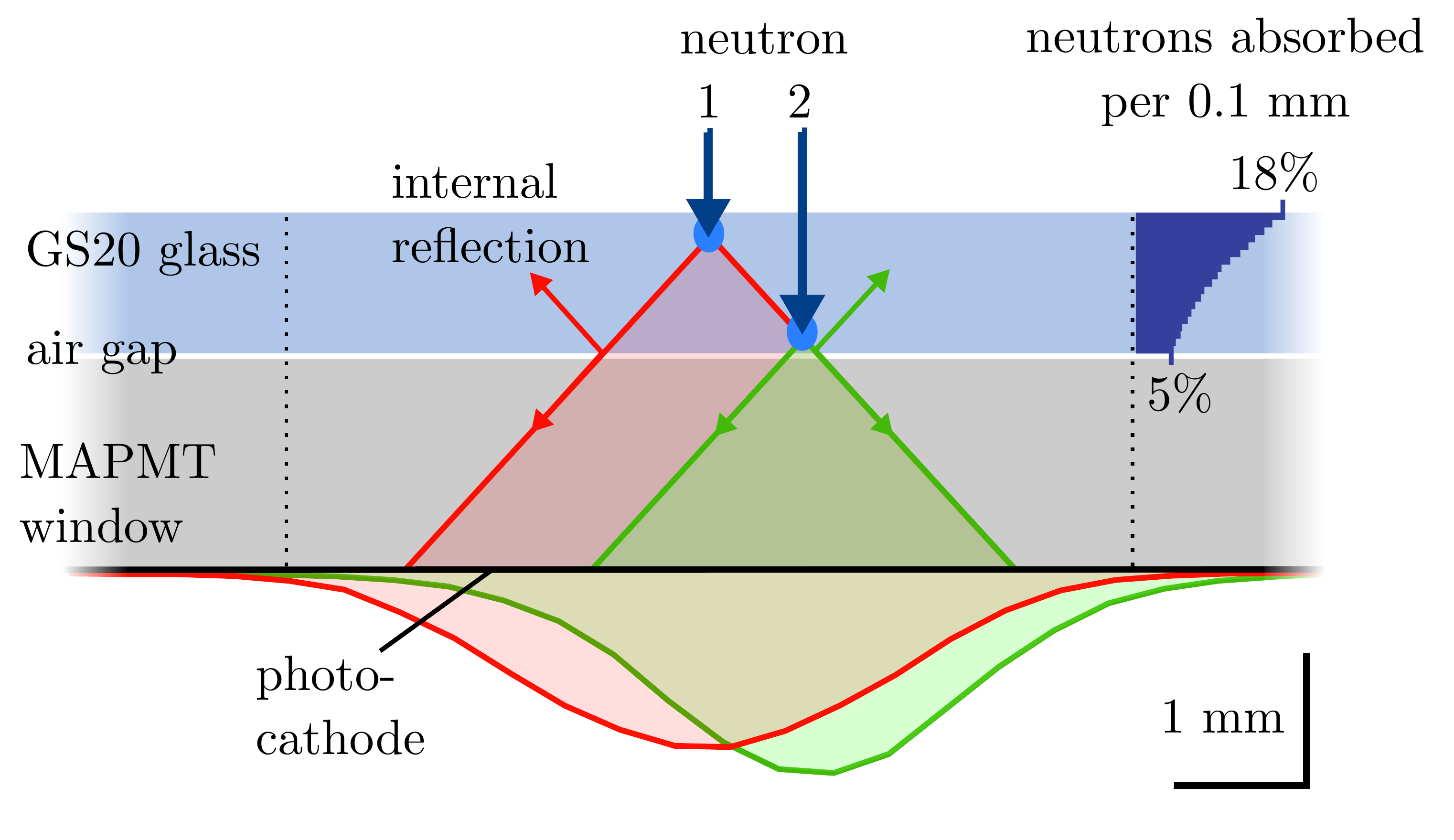}
			\caption{Projected light cones (side view)}
                        \label{subfig:geant4early_late}
		\end{subfigure}
		\begin{subfigure}[b]{0.37\textwidth}
			\includegraphics[width=1\textwidth]{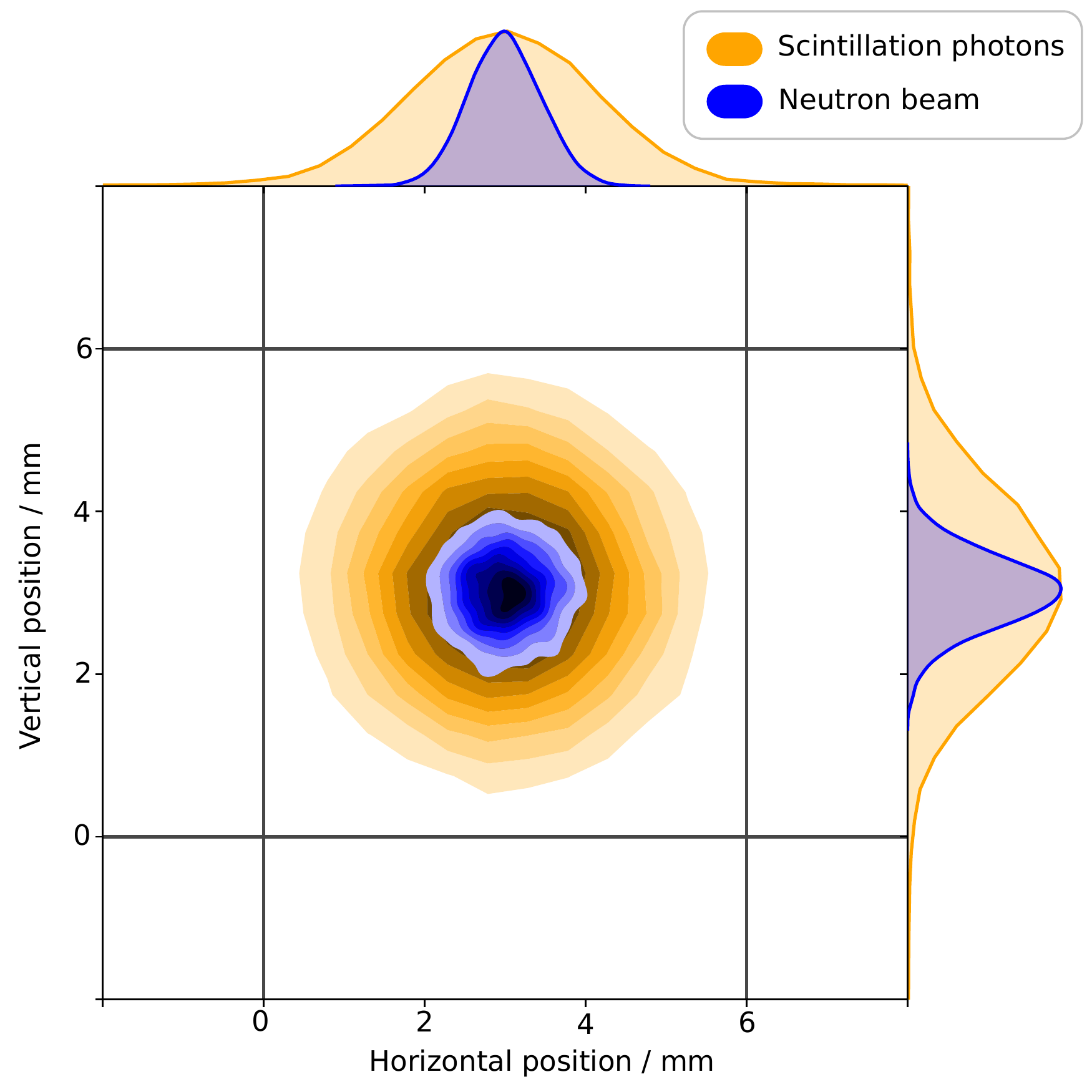}
			\caption{Beam footprints (downstream view)}
			\label{subfig:geant4distributions}
		\end{subfigure}
		\caption{			
			Scintillation-light distributions.
			\ref{subfig:geant4early_late}: 2D projections of 
			\geant~simulated scintillation-light cones (triangles) 
			and yields (inverted normal distributions) resulting 
			from interactions at the upstream GS20 surface 
			(neutron~1, red) and the downstream GS20 surface 
			(neutron~2, green). Internal reflection at the air gap 
			between the scintillator and the MAPMT window is also 
			illustrated. Vertical dotted lines indicate pixel edges. 
			The number of interacting neutrons is shown as a 
			function of penetration depth into the GS20 in the 
			top-right corner.
			\ref{subfig:geant4distributions}: \geant~simulated
			1\,mm FWHM thermal neutron beam (blue) and 
			resulting $\sim$3.5\,mm FWHM scintillation-light 
			distribution (yellow) detected at the MAPMT photocathode 
			for a central irradiation of a pixel. XY projections of 
			the neutron beam and the scintillation-photon 
			distributions are also shown. The vertical and horizontal 
			lines represent the pixel boundaries.
			(For interpretation of the references to color in this 
			figure caption, the reader is referred to the web version 
			of this article.)
			}
		\label{figure:g4}
	\end{center}
\end{figure}

A detailed computer model of a SoNDe module is nearing 
completion~\cite{boyd20}. This C++ model employs the \geant~Monte Carlo 
toolkit~\cite{agostinelli03} version 4.10.6~\cite{allison06}. It includes the GS20 
sheet together with the glass window and photocathode of the MAPMT. Optical-coupling 
media may be placed between the GS20 and the MAPMT window. The model simulates the 
interactions of ionizing radiation in the GS20 to the level of the emission of 
scintillation light and includes the transport of the scintillation photons to the 
MAPMT cathode. It also includes a model for electronic crosstalk. Electronic 
crosstalk results from voltage divider biasing, stray capacitances leading to AC 
coupling between pixel anodes, and charge sharing across neighboring dynode chains, 
all known to affect the performance of the H12700 MAPMT. It results in signal 
from the illuminated pixel leaking into neighboring pixels. Defined for each 
neighboring pixel as the ratio of the induced signal to the signal registered in 
the illuminated pixel, it has been reported to be up to $\sim$3\% in vertically 
adjacent pixels and up to $\sim$7\% in horizontally adjacent pixels~\cite{calvi15}. 
The probability of signal leakage has been shown to be lowest at a pixel center 
and highest at pixel edges. Based on these measurements, electronic crosstalk was 
modelled on an individual scintillation-photon basis with the crosstalk probability
increasing linearly as the pixel edge was approached. For uniform pixel 
illumination, the crosstalk model was configured so that once all scintillation 
photons were detected, adjacent pixels each registered 5\% of the signal 
detected in the illuminated pixel.
Figure~\ref{figure:g4} presents 
some results from the \geant~simulation of the scintillation light. In 
Fig.~\ref{subfig:geant4early_late}, 2D projections of scintillation-light cones 
resulting from individual neutrons interacting at the upstream and downstream 
faces of the GS20 sheet are shown, together with the resulting simulated 
distributions of scintillation light arriving at the photocathode of the MAPMT. 
The projections represent the opening angle defined by total internal 
reflection, and have been drawn to guide the eye.
The scintillation photons from the absorption of a single upstream neutron
have a normal distribution with $\sim$3.5\,mm FWHM at the photocathode, and 
can completely illuminate an entire pixel, even reaching into the adjacent pixels. 
The absorption of a single downstream neutron results in a $\sim$3\,mm 
FWHM normal distribution of scintillation photons at the photocathode. In the 
first 0.1\,mm of the GS20 
sheet, $\sim$18\% of the incoming neutrons are absorbed by the $^{6}$Li in the 
scintillator. This process continues exponentially so that $\sim$5\% of the 
incident neutrons are absorbed in the last 0.1\,mm of the 1\,mm thick GS20.
There is a 22\% chance of a neutron passing through the GS20 without interacting. 
In Fig.~\ref{subfig:geant4distributions}, the relationship between the 1\,mm 
FWHM extended \SI{2.0}{\angstrom} neutron beam incident upon the middle of a 
pixel and the $\sim$3.5\,mm FWHM distribution of scintillation photons at the 
photocathode of a pixel is shown. The 3\,mm FWHM neutron-beam scintillation-light  
footprint covers essentially the entire MAPMT pixel.
The paper shown in Fig.~\ref{subfig:schematic} diffusely scatters the 
scintillation light back towards the photocathode, resulting in a $\sim$40\% 
increase in the detected yield, and increasing the width of the light cone 
by $\sim$2\%.

\section{Measurement}
\label{section:measurement}

Collimated thermal-neutron beams were used to 
irradiate the SoNDe module at well-defined positions. After passing through 
the hole in the Mirrobor sheet, neutrons entered the black box which was 
positioned on an XY coordinate scanner instrumented with two translation stages 
(M-IMS600 and M-IMS300) and a motor controller (ESP301), all from MKS Newport
Corporation~\cite{newport}. The SoNDe module was located inside 
the black box and positioned so that its face was parallel to the upstream 
side of the black box, and both were perpendicular to the neutron beam. The 
beam struck the upstream face of the GS20 sheet after passing through a thin 
layer of tape and and white paper. The SoNDe module was stepped through the 
neutron beams with a stepsize of 0.5-1\,mm in the X and Y~directions. The anode 
signals from each of the MAPMT pixels were processed using the dedicated SoNDe 
electronics. Negative polarity analog pulse heights for each event with at least 
one pixel producing a signal above the 500~ADC channel threshold were recorded. 
Data were 
recorded for $\sim$15~s (10000~events) at each point on a scan, 
followed by a motor translation, so that a complete scan of 2~$\times$~2 pixels 
with 0.5\,mm spacing took several hours. 
Analysis of the data was performed using \python-based~\cite{python} 
\pandas~\cite{pandas} and \SciPy~\cite{mckinney10} tools. 

\section{Results}
\label{section:results} 

\begin{figure}[H]
	\begin{center}
		\includegraphics[width=0.6\textwidth]{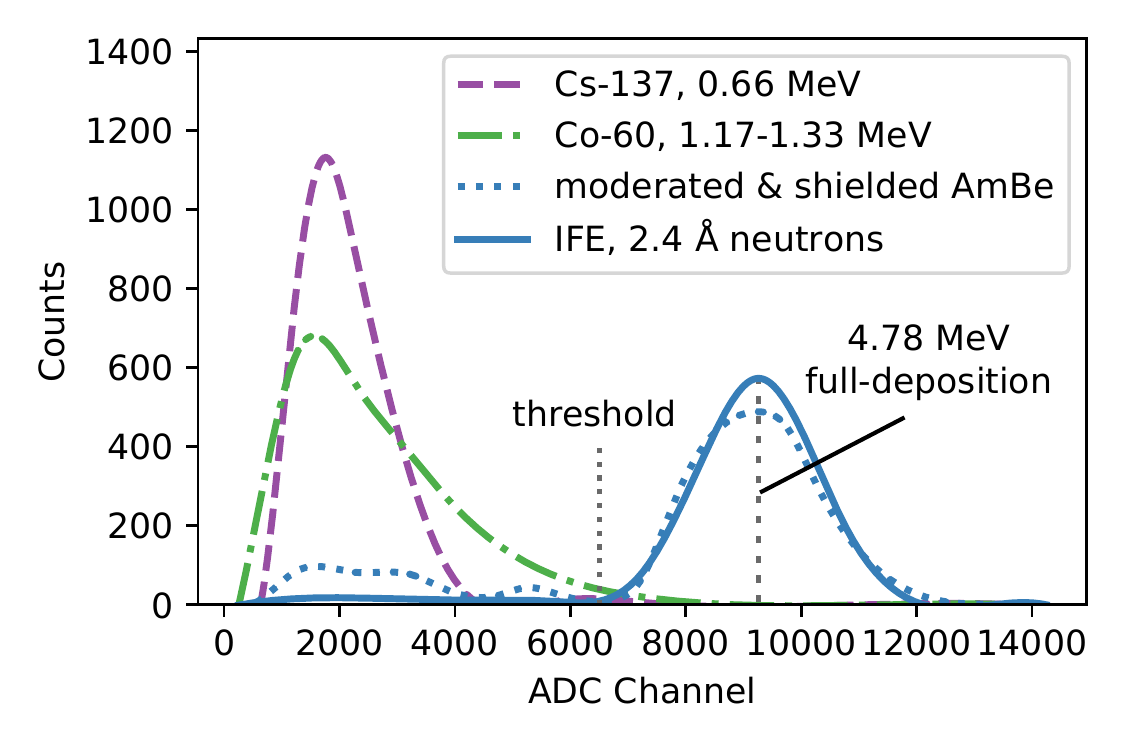}
		\caption{
			$\gamma$-ray and neutron calibration spectra for P37. 
			Events resulting in the largest signal in P37 are 
			displayed. Spectra from sources of $\gamma$-rays (long 
			purple 
			dashes, green dash-dots) lie to the left (ADC 
			channel~3000) while spectra from neutrons (solid 
			blue, blue dots) lie to the right 
			(ADC channel~9090). A typical threshold is indicated.
			(For interpretation of the references to color in this
			figure caption, the reader is referred to the web 
			version of this article.)
		}
		\label{figure:sources}
	\end{center}
\end{figure}

Figure~\ref{figure:sources} shows results from irradiations of the detector 
with uncollimated $^{60}$Co (E$_{\gamma}$~$=$~1.17, 1.33\,MeV) and $^{137}$Cs 
(E$_{\gamma}$~$=$~662\,keV) $\gamma$-ray sources, a moderated and heavily
$\gamma$-ray shielded (but uncollimated) Am/Be neutron source, and the 3\,mm, 
\SI{2.4}{\angstrom} neutron beam directed at the center of P37. Note that none 
of the source irradiations occurred in situ, but instead were performed 
subsequently with the same experimental equipment and setup parameters at the
Source-Testing Facility at the University of Lund in Sweden~\cite{messi17}. 
For each irradiation, 
events corresponding to the largest signal in P37 were selected. The 4.78\,MeV 
full-deposition peak resulting from neutron capture on $^{6}$Li ($\sim$6600 
scintillation photons) is located at about ADC channel 9090 
($\sim$0.53\,keV/channel). The distribution 
from $\sim$1\,MeV $\gamma$-rays, which are typical backgrounds at accelerator 
facilities such as ESS, slightly overlap the neutron peak. A discriminator 
threshold of $\sim$72\% of the mean of the full-deposition peak (ADC channel 
6500) discriminates against $\sim$93\% of the detected $\sim$1\,MeV 
$\gamma$-rays while retaining the neutron peak.

\begin{figure}[H]
	\begin{center}
		\begin{subfigure}[b]{0.45\textwidth}
			\includegraphics[width=1.\textwidth]{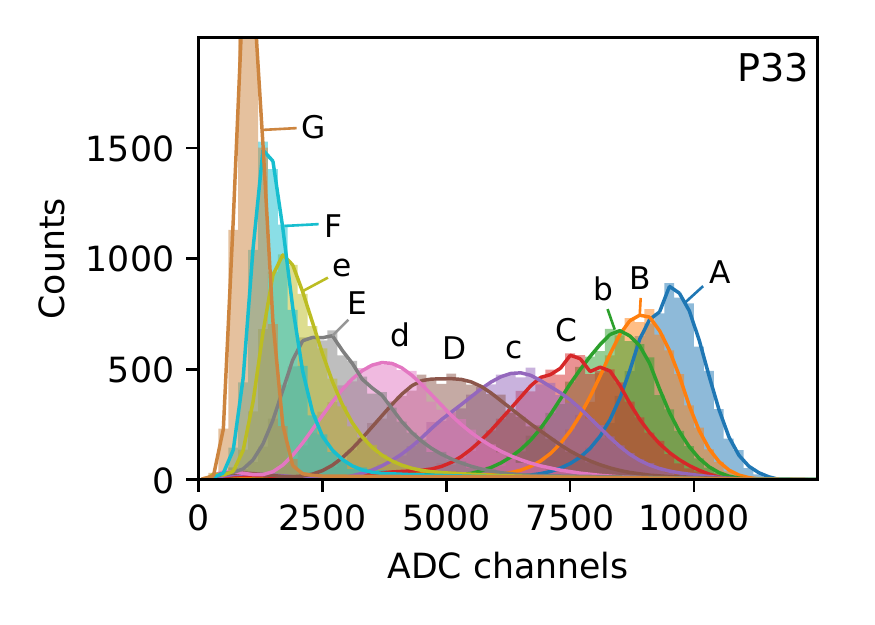}
			\caption{P33, collected charge}
			\label{subfig:neut_hor_P33}
		\end{subfigure}
		\begin{subfigure}[b]{0.45\textwidth}
			\includegraphics[width=1.\textwidth]{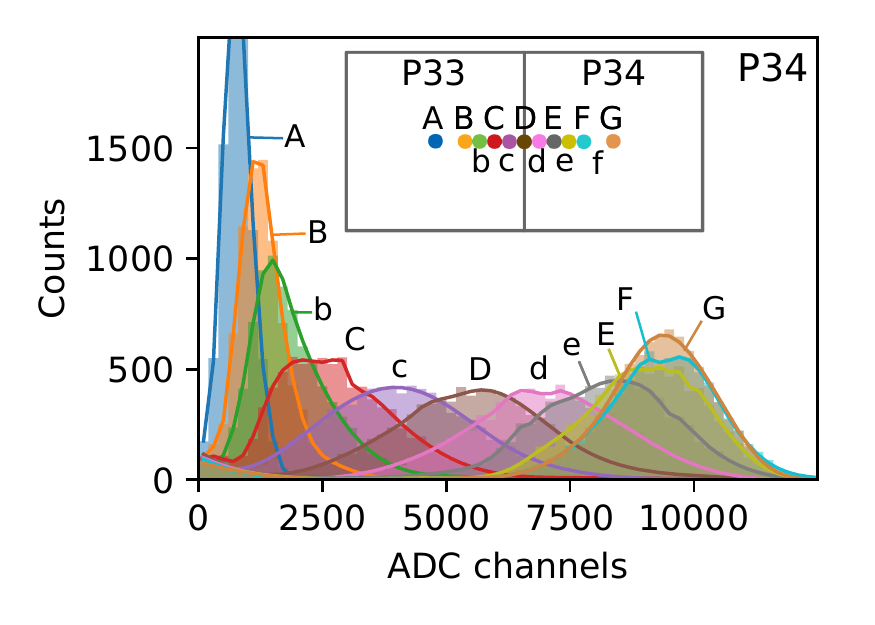}
			\caption{P34, collected charge}
			\label{subfig:neut_hor_P34}
		\end{subfigure}
		\begin{subfigure}[b]{0.45\textwidth}
			\includegraphics[width=1.\textwidth]{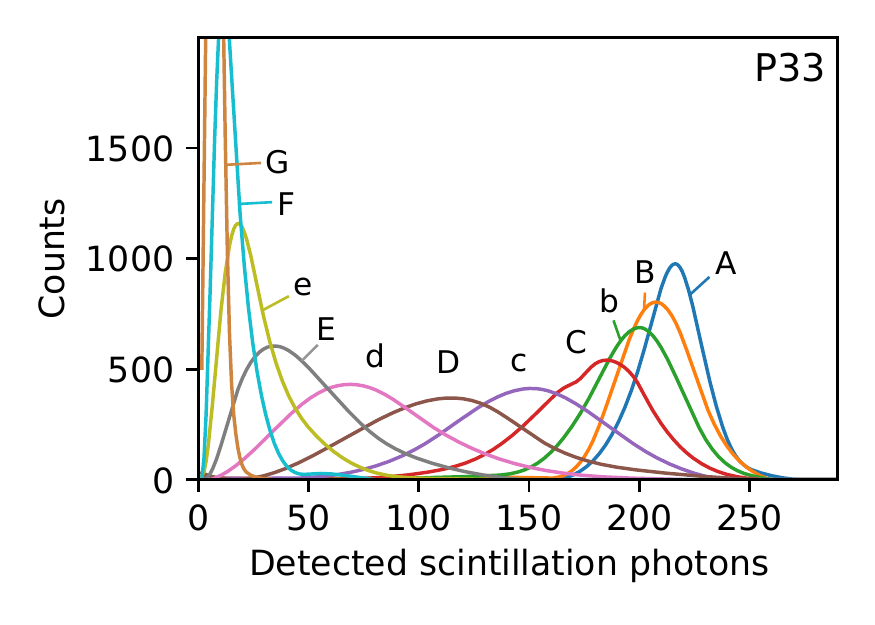}
			\caption{P33, \geant}
			\label{subfig:neut_hor_sim_P33}
		\end{subfigure}
		\begin{subfigure}[b]{0.45\textwidth}
			\includegraphics[width=1.\textwidth]{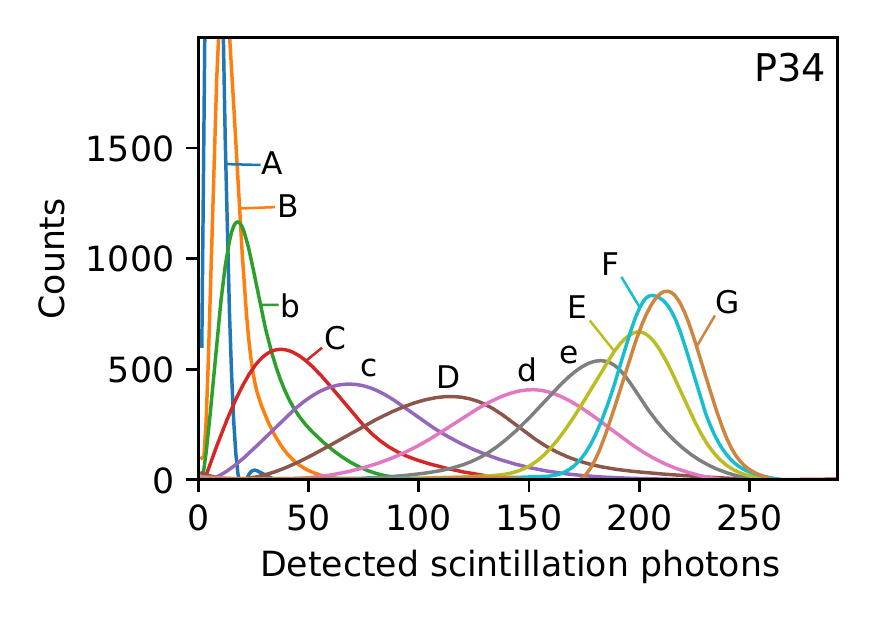}
			\caption{P34, \geant}
			\label{subfig:neut_hor_sim_P34}
		\end{subfigure}
		\caption{
			Scintillation-light sharing. Horizontal scan from P33 to P34 
			using the 1\,mm diameter FWHM \SI{2.0}{\angstrom} neutron 
			beam in 0.5\,mm steps. The colors and beam 
			locations are presented in the key (inset, top right 
			panel) which applies to both the measured gain-corrected 
			charge distributions (filled histograms, top panels) and 
			the \geant~simulations of the scintillation-light yields 
			(open histograms, bottom panels). Spectra for positions 
			a and f are not shown as they are almost identical to 
			the spectra obtained at the adjacent positions. 
			(For interpretation of the references to color in this 
			figure caption, the reader is referred to the web version 
			of this article.) 
			}
		\label{figure:neut_hor}
	\end{center}
\end{figure}

As previous 
work~\cite{korpar00,rielage01,matsumoto04,lang05,abbon08,montgomery12,montgomery15,rofors19,rofors20}
has clearly demonstrated that MAPMT pixel-gain maps depend strongly upon the 
method of illumination, all of the results presented below have been pedestal 
subtracted and gain corrected with pixel-gain maps produced from 3\,mm, 
\SI{2.4}{\angstrom} neutron-beam irradiations of the pixel centers.
Figure~\ref{figure:neut_hor} shows data and \geant~simulations for a horizontal
scan of the SoNDe module through the 1\,mm FWHM, $\sim$\SI{2.0}{\angstrom} neutron 
beam. The module was moved from position~A (center, edge pixel P33) to position~G 
(center, P34) in 0.5\,mm steps. For 13 scan positions, neutron pulse-height spectra 
and \geant-simulated scintillation-light yields are shown for P33 
(Figs.~\ref{subfig:neut_hor_P33}, \ref{subfig:neut_hor_sim_P33}) and P34 
(Figs.~\ref{subfig:neut_hor_P34}, \ref{subfig:neut_hor_sim_P34}). The simulations
include the nominal 5\% electronic crosstalk contribution previously discussed
(see also the discussion associated with 
Fig.~\ref{figure:mean_yields_with_gain_correction} below).
The location of the neutron beam determines the amount of scintillation light 
collected by a single pixel. The single-pixel signal amplitude is largest when the 
neutron beam strikes the pixel center, smaller when the neutron beam strikes the 
pixel edge, and smallest when the neutron beam strikes the pixel corner. 
Scintillation produced when the beam strikes the border between two pixels is 
equally shared, which results in equal signal amplitudes after pedestal and gain 
correction. The excellent agreement between the data and the simulation 
indicates that both the scintillation-light sharing and the resulting pixel response 
are well-understood.

Figure~\ref{figure:mean_yields_with_gain_correction} illustrates 
scintillation-light sharing by edge pixel P33 and adjacent P34 as the SoNDe 
module was translated horizontally through the 1\,mm FWHM,
$\sim$\SI{2.0}{\angstrom} neutron beam. Figure~\ref{subfig:lightshare_neut}
shows the means of 
the pulse-height distributions (from Fig.~\ref{figure:neut_hor}) obtained for 
each beam position. The scan shows that signal 
leakage to adjacent pixels is $\sim$7-12\% when the neutron beam strikes the 
center of either pixel. This represents a larger spread of scintillation signal 
into the adjacent pixel than was the case for previous investigations of 
relatively central pixels with charged-particle beams~\cite{rofors19,rofors20}
and may be related to the diffusely reflecting white paper placed at the front 
face of the GS20 sheet. $\alpha$-particle scan results for (non-edge) P36, P37, 
P44, and P45~\cite{rofors19} demonstrated summed gain-corrected charge 
distributions that were flat across the pixels and boundary regions.
Proton- and deuteron-scan results for (non-edge) P37 and P38~\cite{rofors20} 
demonstrated summed gain-corrected charge distributions that were slightly 
convex and centered at 
the pixel edge. This was because the pixels together collected slightly more of 
the scintillation light produced from an event at the boundary between them 
than they collected from an event at the center of either pixel, with the 
missing light collected by the surrounding pixels. Here, the measured 
distributions may indicate a light-collection enhancement when P34 is 
irradiated. Simulations including the nominal 5\% level of electronic crosstalk 
underestimate the amount of signal leaking into the adjacent pixel. 
Crosstalk measurements~\cite{calvi15} suggest that 5\% is already an overestimate.
Figure~\ref{subfig:lightshare_normalized} shows the light-sharing 
ratio between P33 and P34 defined as (P33$-$P34)/(P33+P34). For the nominal 
5\% level of electronic crosstalk, the simulation results in too much signal 
in the irradiated pixel relative to the adjacent pixel. Agreement at the border 
between pixels is very good. Very recent simulations~\cite{boyd20} indicate that 
the matt white reflector (paper) upstream of the scintillator and the glass 
surface finish will likely contribute to the redistribution of scintillation 
light in a fashion similar to electronic crosstalk. Unfolding these effects will 
require a detailed study with a laser.

\begin{figure}[H]
	\begin{center}
		\begin{subfigure}[b]{0.5\textwidth}
			\includegraphics[width=1.\textwidth]{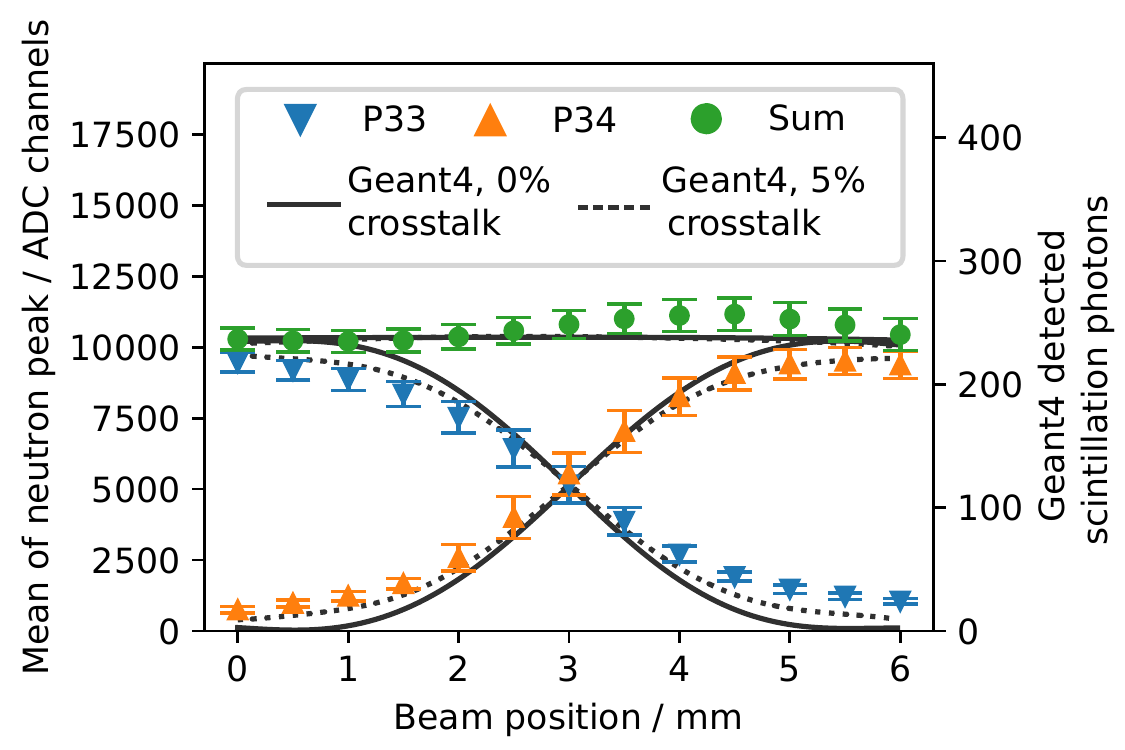}
			\caption{Light share, neutron beam}
			\label{subfig:lightshare_neut}
		\end{subfigure}
		\begin{subfigure}[b]{0.44\textwidth}
			\includegraphics[width=1\textwidth]{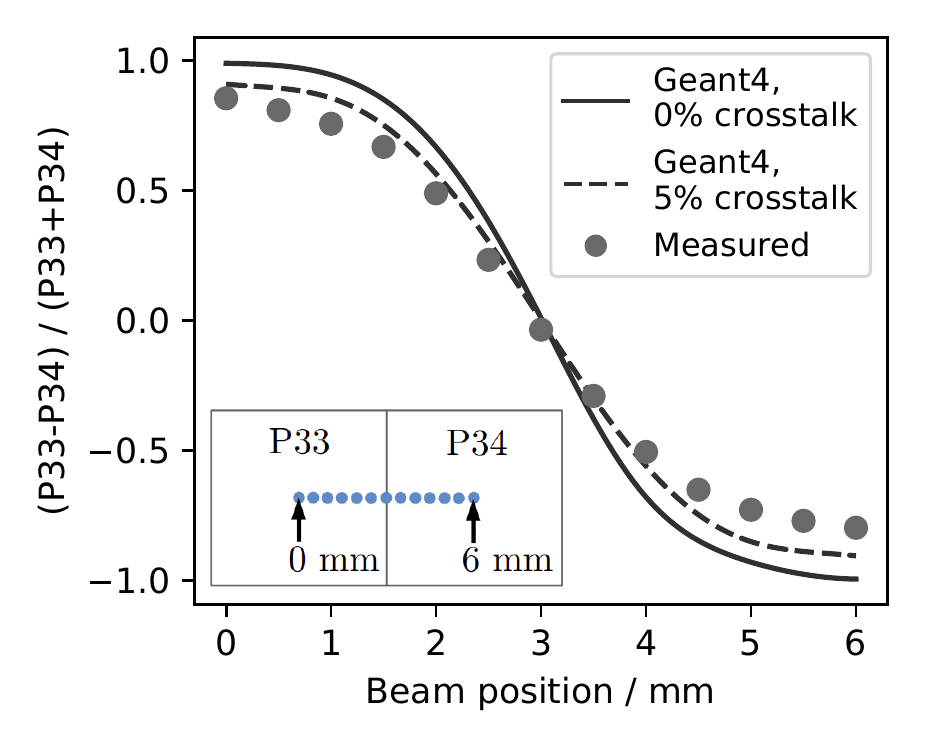}
			\caption{Light share, normalized}
			\label{subfig:lightshare_normalized}
		\end{subfigure}
		\caption{
			Scintillation-light sharing. Horizontal scan from P33 to 
			P34 through the a 1\,mm diameter FWHM \SI{2.0}{\angstrom} 
			neutron beam in 0.5\,mm steps.
			\ref{subfig:lightshare_neut}: Data and \geant~simulations. 
			Points are the means from fits to the distributions of 
			scintillation light registered in the pixels shown in 
			Fig.~\ref{figure:neut_hor}. The curves are spline fits to 
			the corresponding \geant~simulations. The 
			uncertainties in the means are smaller than the symbols.
			The error bars correspond to $\pm$~$\sigma/4$ of the 
			fitted distributions. The curves result from the 
			\geant~simulations of the scintillation light, and 
			have been normalized to the measurements as previously
			discussed. The curves show the simulations including different
			levels of electronic crosstalk between pixels.
			\ref{subfig:lightshare_normalized}: Light-sharing ratio 
			derived from \ref{subfig:lightshare_neut}. The uncertainies 
			are smaller than the widths of the lines. The curves are as 
			in Fig.~\ref{subfig:lightshare_neut}.
			}
		\label{figure:mean_yields_with_gain_correction}
	\end{center}
\end{figure}

\begin{figure}[H]
	\centering
		\begin{subfigure}[b]{0.6\textwidth}
			\includegraphics[width=1\textwidth]{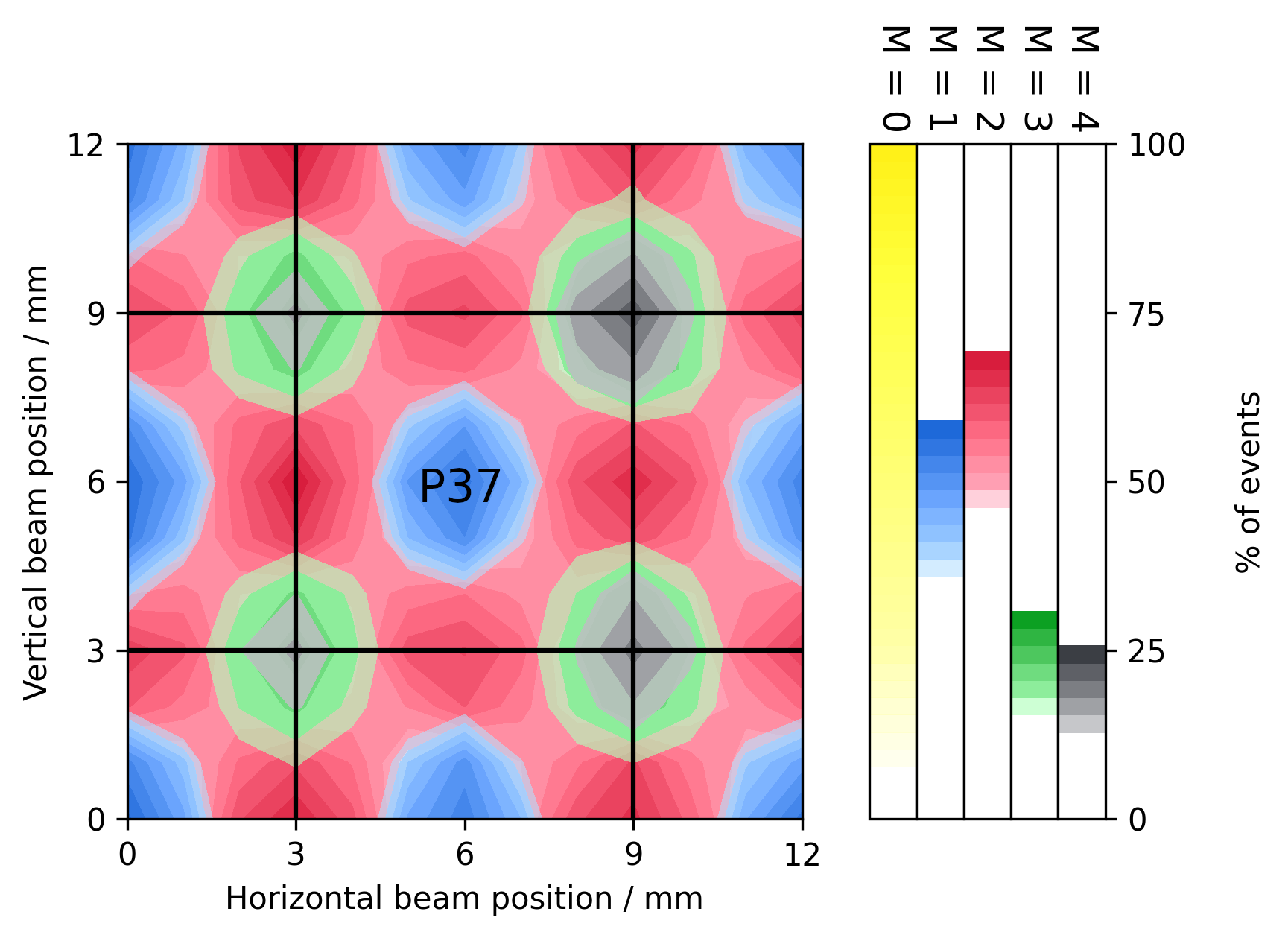}
			\caption{Spatial, threshold 15\% of full peak (1360 ADC~channels)}
			\label{subfig:1360_channels}
		\end{subfigure}
		\begin{subfigure}[b]{0.6\textwidth}
			\includegraphics[width=1\textwidth]{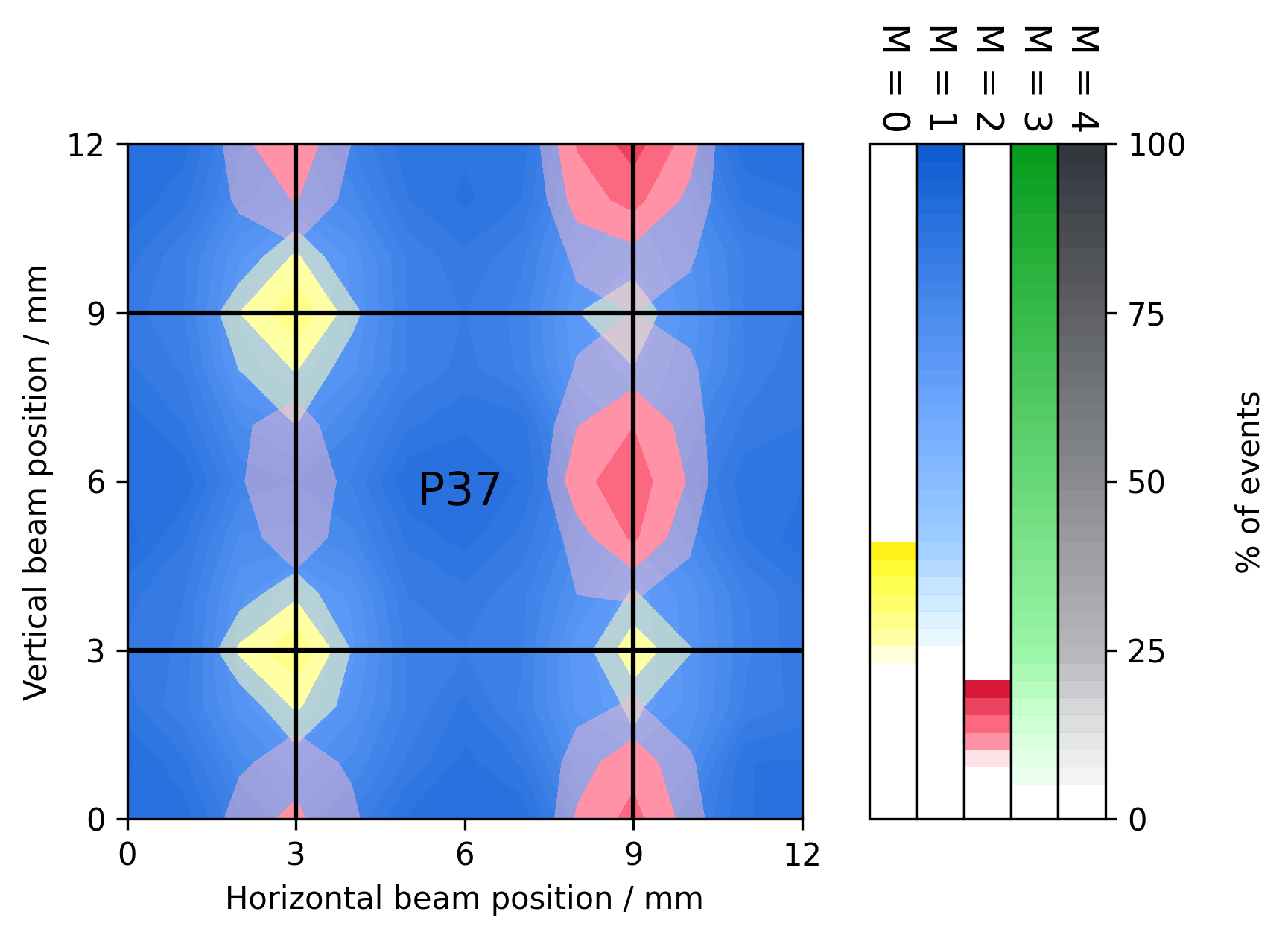}
			\caption{Spatial, threshold 50\% of full peak (4545 ADC~channels)}
			\label{subfig:4545_channels}
		\end{subfigure}
		\begin{subfigure}[b]{0.55\textwidth}
			\includegraphics[width=1\textwidth]{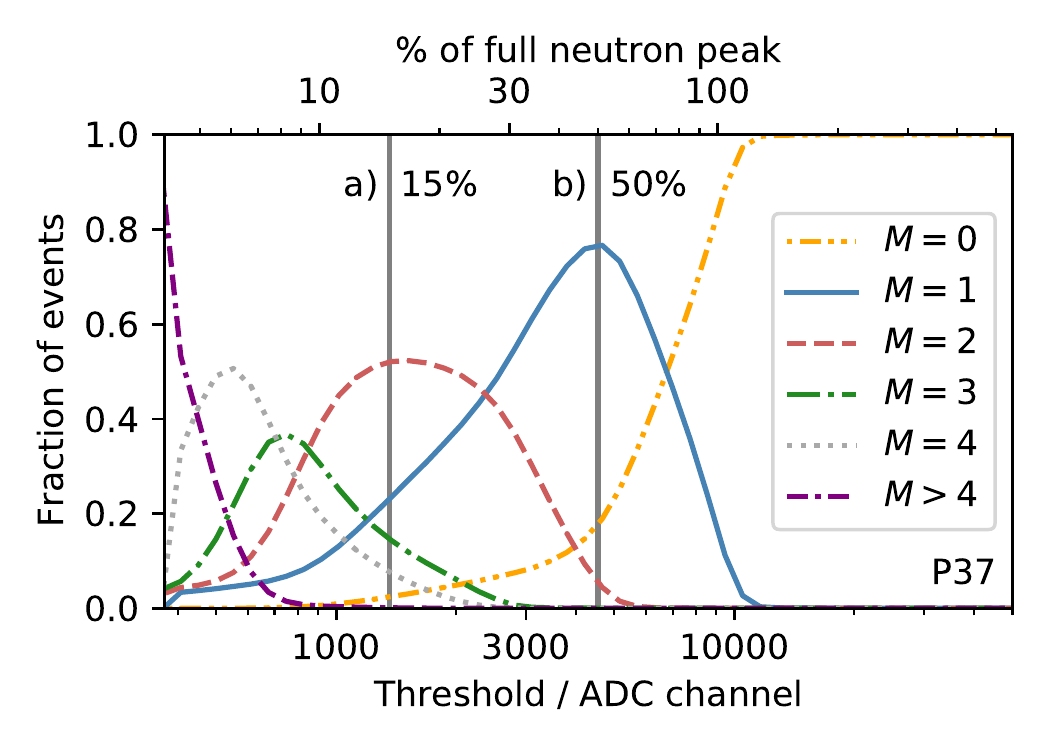}
			\vspace*{-8mm}
			\caption{Fractional, various thresholds}
			\label{subfig:mult_per_threshold}
		\end{subfigure}
		\caption{Multiplicity distributions for the 1\,mm FWHM
			\SI{2.0}{\angstrom} beam incident on P37 and the 
			surrounding pixels. In \ref{subfig:1360_channels} 
			(threshold 1360~ADC channels, 15\% of the 
			full-deposition peak) and \ref{subfig:4545_channels} 
			(threshold 4545~ADC channels (50\% of the full-deposition 
			peak), the black lines denote the pixel boundaries. 
			Yellow indicates $M$~$=$~0 events, blue indicates 
			$M$~$=$~1 events, red indicates $M$~$=$~2 
			events, green indicates $M$~$=$~3 events, and grey
			indicates $M$~$=$~4 events. The lighter the shade of the 
			color, the fewer the number of events. 
			\ref{subfig:mult_per_threshold} presents the fraction 
			of events registered in P37 for each multiplicity as 
			a function of threshold common to all pixels, with 
			the 1360 and 4545~ADC channel thresholds shown as 
			vertical lines. 
			(For interpretation of the references to color in this
			figure caption, the reader is referred to the web 
			version of this article.)
			}
		\label{figure:contour_maps_M_with_cuts}
\end{figure}

The pixel-hit multiplicity ($M$~$=$~1, $M$~$=$~2, etc.) for events as a 
function of the beam-spot position has previously been studied with scans of 
$\sim$1\,mm FWHM beams of $\alpha$-particles~\cite{rofors19} and 
$\sim$\SI{100}{\micro\meter} diameter beams of protons and 
deuterons~\cite{rofors20}. A hit occurred if a pixel amplitude exceeded a 
threshold which was varied offline. Here, the 1\,mm FWHM \SI{2.0}{\angstrom} neutron 
beam was employed in a complementary study. Neutron-beam irradiations with a 
stepsize of 1\,mm in X and Y were performed resulting in a 13~$\times$~13 matrix 
of data. Figure~\ref{figure:contour_maps_M_with_cuts} displays 2D position dependence 
of tions obtained near P37 for software thresholds of 1360 
(Fig.~\ref{subfig:1360_channels}) and 4545 (Fig.~\ref{subfig:4545_channels}) 
ADC channels, which correspond to $\sim$15\% and $\sim$50\% of the mean of the 
P37 pixel-centered full-deposition neutron peak (4.78\,MeV, 0.52\,keV/channel), 
respectively. For the 1360 ADC channel threshold, $\sim$2\% of events are lost, and 
$M$~$=$~1 events ($\sim$22\%) 
lie within $\pm$1\,mm of the pixel center. The $M$~$=$~2, 3, and 4 data are all 
localized to $\sim$5\,mm, within the 6\,mm position-resolution constraint for SoNDe 
operation at ESS. A threshold of 4545 ADC channels results in $\sim$18\% event loss,
maximizes both the number of $M$~$=$~1 events ($\sim$78\%) and the area of the 
detector where $M$~$=$~1 events may be detected.
Figure~\ref{subfig:mult_per_threshold} shows event multiplicity as a function of
applied threshold for $M$~$=$~1$-$4. The curves all have well-defined maxima so 
that the multiplicity $M$ for a dataset can be selected by enforcing the 
appropriate threshold. For example, for a dataset of $\sim$78\% $M$~$=$~1, $\sim$4\% 
$M$~$=$~2, and a negligible number of $M$~$=$~3, 4 events, a threshold of 4545~ADC 
channels must be applied. A result of operating the SoNDe module with this 
relatively high threshold is that $\sim$18\% of events have $M$~$=$~0 and are thus 
lost.

\begin{figure}[H]
	\centering
		\begin{subfigure}[b]{0.45\textwidth}
			\includegraphics[width=1\textwidth]{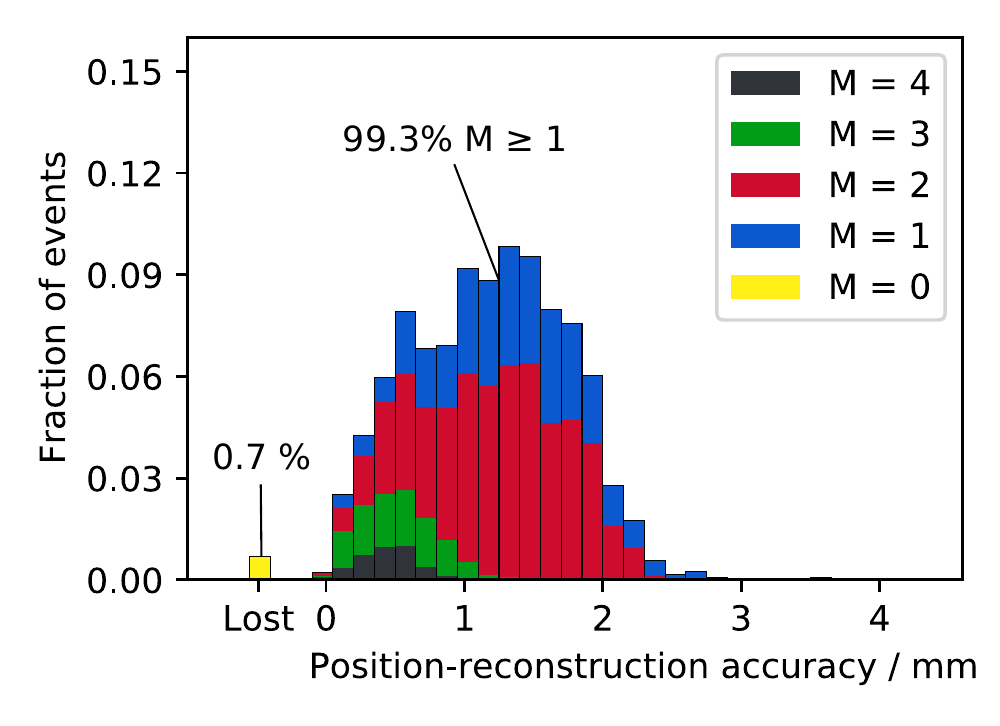}
			\caption{Threshold 15\% of full peak (33 scintillation photons)}
			\label{subfig:posres15}
		\end{subfigure}
		\begin{subfigure}[b]{0.45\textwidth}
			\includegraphics[width=1\textwidth]{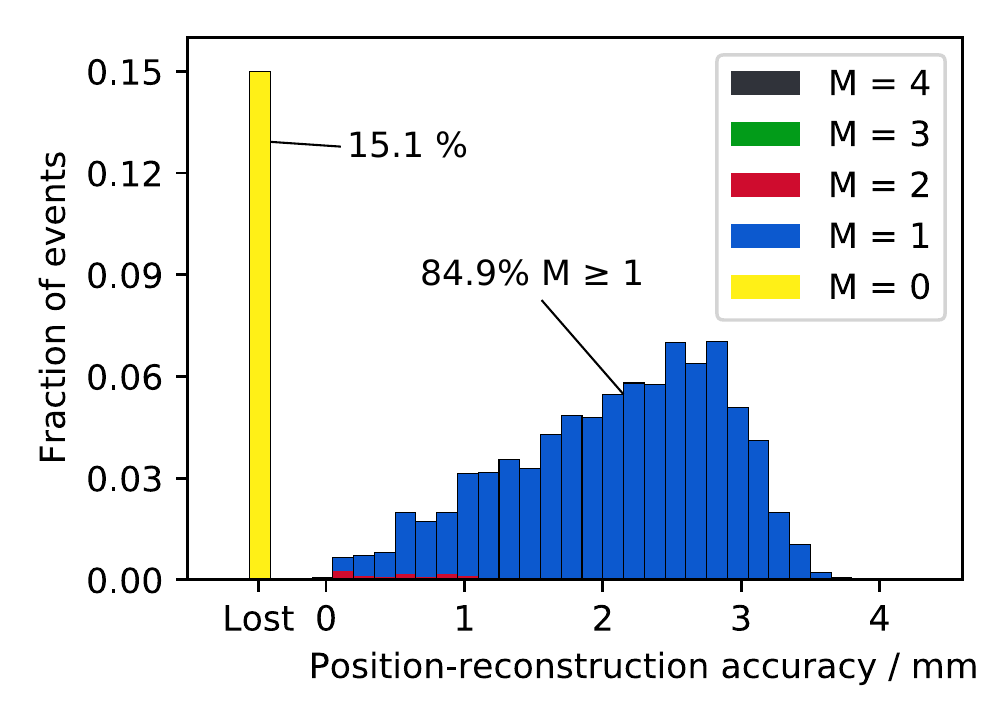}
			\caption{Threshold 50\% of full peak (100 scintillation photons)}
			\label{subfig:posres50}
		\end{subfigure}
		\caption{
			\geant~simulations of the position-reconstruction accuracy 
			for \SI{2.0}{\angstrom} neutrons incident on P37 and the 
			surrounding pixels. In \ref{subfig:posres15}
			(threshold 33 scintillation photons, 15\% of the 
			full-deposition peak) and \ref{subfig:posres50} (threshold
			100 scintillation photons, 50\% of the full-deposition
			peak), the color code is the same as in 
			Fig.~\ref{figure:contour_maps_M_with_cuts}.
			(For interpretation of the references to color in this
			figure caption, the reader is referred to the web 
			version of this article.)
			}
		\label{figure:position_resolution}
\end{figure}

Figure~\ref{figure:position_resolution} shows the \geant-simulated 
position-reconstruction accuracy as a function of pixel-hit multiplicity 
for a 15\% threshold corresponding to 33 scintillation photons 
(Fig.~\ref{subfig:posres15}) and a 50\% threshold corresponding
to 100 scintillation photons (Fig.~\ref{subfig:posres50}). The accuracy
is defined as the distance between the reconstructed and simulated capture
vertex. The area in the vicinty of P37 was uniformly illuminated with 
\SI{2.0}{\angstrom} neutrons. For each event, the reconstructed hit position 
was given by the average of the locations of the pixel centers of all pixels 
registering a signal above threshold. This method results in $M$~$=$~1 events 
being ``assigned" a position in the center of the pixel above threshold, while 
$M=2$ events are positioned at the edge between the two pixels above 
threshold. $M=3$ events are positioned within the triangle defined by the 
pixel centers, offset 1\,mm in both X and Y from the common corner, and $M=4$ 
events are positioned at the common corner. For the 33 scintillation-photon 
threshold, $\sim$1\% of events are lost. Event positions for multiplicities 
$M$~$=$~1 ($\sim$31\%) and $M$~$=$~2 ($\sim$56\%) events are reconstructed to 
better than $\sim$3\,mm, while $M$~$=$~3 ($\sim$9\%) and $M$~$=$~4 ($\sim$4\%) 
events are reconstructed to better than $\sim$1\,mm. For the 100 
scintillation-photon threshold, $\sim$15\% of events are lost. $M$~$=$~1 
($\sim$84\%) events are reconstructed to better than $\sim$4\,mm, and $M$~$=$~2 
($\sim$1\%) events are reconstructed to better than $\sim$1\,mm. While 
agreement between the simulation and the data is presently not perfect, it is 
very encouraging, and development continues.

\section{Summary and Discussion}
\label{section:summary}

Collimated beams of 13\,meV and 18\,meV neutrons from the IFE reactor
(Fig.~\ref{figure:IFE}) have been used to investigate the position-dependent 
response of a pixelated neutron detector known as a SoNDe module 
(Fig.~\ref{figure:SoNDeModule}). A 
SoNDe module consists of a 1\,mm thick sheet of GS20 scintillating glass coupled 
to a 64~pixel H12700A MAPMT with dedicated readout electronics. The software
layer of the data acquisition (EFU) was a prototype under development for ESS.
The amplitudes of the pixel signals were investigated for different irradiation 
positions by scanning the module through the beam in steps of 0.5-1\,mm using a 
motor-driven XY table. A \geant~model of the SoNDe module
greatly aided in the interpretation of the data (Fig.~\ref{figure:g4}). The 
amount of scintillation light detected by the MAPMT was increased by $\sim$40\% by
placing a sheet of diffusely reflecting white paper at the front face of the GS20.
$\gamma$-rays and neutrons could generally be discriminated with a simple
threshold cut (Fig.~\ref{figure:sources}). The amplitudes of the gain-corrected 
signals were highly dependent on where the neutron beam struck the detector 
(Fig.~\ref{figure:neut_hor}). When directed towards a central-pixel region, 
$\sim$5\% of the signal was detected in an adjacent pixel. However, within 
$\sim$1\,mm of the boundary, $\sim$30\% of the signal was registered in the 
adjacent pixel. At the boundary, the signal was evenly split between pixels. 
Overall agreement between the data and \geant~simulations was good when a 5\% 
level of interpixel electronic crosstalk was considered. The signal in a pixel 
adjacent to an edge pixel when the edge pixel was irradiated was underestimated 
(Fig.~\ref{figure:mean_yields_with_gain_correction}). 
For different beam positions, the effect of raising the pixel threshold on 
the hit multiplicity was studied (Fig.~\ref{figure:contour_maps_M_with_cuts}). 
When the threshold was set at $\sim$50\% of the mean of the neutron 
full-deposition peak, $\sim$$78$\% of the data had $M$~$=$~1, $\sim$4\% were 
$M$~$=$~2, and $\sim$18\% were undetected. Increasing the threshold to higher 
values resulted in $M$~$=$~1 event loss and a reduction of the sensitive area 
of the detector. Decreasing the 
threshold to $\sim$15\% of the mean of the neutron full-deposition peak resulted 
in $\sim$2\% event loss, $\sim$22\% $M$~$=$~1 data, and $\sim$66\% $M$~$>$~1 data. 
The \geant~simulation was employed to investigate the position-reconstruction 
accuracy (Fig.~\ref{figure:position_resolution}) of the measured multiplicity 
regions shown in Fig.~\ref{figure:contour_maps_M_with_cuts}. For the threshold 
set at $\sim$50\% of the mean of the neutron full-deposition peak,
the majority of the events registered with $M$~$=$~1 could be reconstructed to 
better than 3.5\,mm. The majority of the events registered with $M$~$=$~2 could be 
reconstructed to better than 1\,mm. For the threshold set at 
$\sim$15\% of the mean of the neutron full-deposition peak, the majority of the 
$M$~$=$~1 and $M$~$=$~2 events could be recontructed to better than 2.5\,mm, and
the majority of the $M$~$=$~3 and $M$~$=$~4 events could be reconstructed to 
better than 1\,mm.
Thus, all the IFE data could be reconstructed with an accuracy better than the 6\,mm
position resolution required for the operation of SoNDe at ESS.

\section*{Acknowledgements}
\label{acknowledgements}

Support for this project was provided by the European Union via the Horizon 
2020 Solid-State Neutron Detector Project (Proposal ID 654124) and the 
BrightnESS Project (Proposal ID 676548). Support was also provided by the
UK Science and Technology Facilities Council (Grant No. ST/P004458/1) and 
the UK Engineering and Physical Sciences Research Council Centre for Doctoral 
Training in Intelligent Sensing and Measurement (Grant No. EP/L016753/1).
The authors thank the staff of IFE for providing the beams of neutrons.

\newpage

\bibliographystyle{elsarticle-num}

\end{document}